\documentclass[aps, showpacs, twocolumn, amsmath, amssymb, pra]{revtex4}
\usepackage{graphicx} 
\usepackage{color}
\usepackage[T1]{fontenc}
\usepackage{ae,aecompl}
\usepackage{pstricks}
\usepackage{pstricks-add}

\newcommand{\ci}{c_{i}^{\phantom{\dagger}}}

\newcommand{\cj}{c_{j}^{\phantom{\dagger}}}
\newcommand{\cjd}{c_{j}^{\dagger}}

\newcommand{\ck}{c_{k}^{\phantom{\dagger}}}
\newcommand{\ckd}{c_{k}^{\dagger}}

\newcommand{\cmk}{c_{-k}^{\phantom{\dagger}}}
\newcommand{\cmkd}{c_{-k}^{\dagger}}

\newcommand{\gk}{\gamma_{k}^{\phantom{\dagger}}}
\newcommand{\gkd}{\gamma_{k}^{\dagger}}

\newcommand{\cnoll}{c_{0}^{\phantom{\dagger}}}
\newcommand{\cnolld}{c_{0}^{\dagger}}
\newcommand{\cpi}{c_{\pi}^{\phantom{\dagger}}}
\newcommand{\cpid}{c_{\pi}^{\dagger}}
\newcommand{\gnoll}{\gamma_{0}^{\phantom{\dagger}}}

\begin{document}


\title{Multicriticality and entanglement in the one-dimensional quantum compass model}


\author{Erik Eriksson}
\author{Henrik Johannesson}
\affiliation{Department of Physics, University of Gothenburg, SE 412 96 Gothenburg, Sweden}


\date{\today}

\begin{abstract}
We study the one-dimensional (1D) quantum compass model with two
independent parameters by means of an exact mapping to the
quantum Ising model. This allows us to uncover hidden features of the
quantum phase transition in the ordinary one-parameter quantum compass model, 
showing that it occurs at a multicritical point where a line of first-order transitions intersects a line of second-order symmetry-breaking transitions of Ising type. We calculate the concurrence 
and the block entanglement entropy in the four ground state phases, and find that these entanglement measures accurately signal the second-order, but not the first-order, transitions.
\end{abstract}


\maketitle


\section{Introduction}
The orbital degeneracy in {\em d}-shell transition metal oxides underpins many of the fascinating phenomena found in these materials.  Well-known examples include ferroelectricity, colossal magnetoresistance, and charge ordering \cite{Cheong}. An early attempt to model the directional nature 
of the orbital states in the case of a two-fold degeneracy was made by Kugel and Khomskii \cite{KugelKhomskii}, who introduced a simplified model $-$ the {\em quantum compass model} (QCM) $-$  where the orbital degrees of freedom are represented by (pseudo)spin-1/2 operators and coupled anisotropically in such a way as to mimic the competition between orbital orderings in different 
directions. For example, on a two-dimensional (2D) square lattice, the coupling between spin components on neighboring lattice sites is tied to the direction of the corresponding bonds, with Ising-like interactions $J_x\sigma_i^x\sigma_{i+\hat{x}}^x$ and $J_y\sigma_i^y\sigma_{i+\hat{y}}^y$ along the $\hat{x}$- and $\hat{y}$-axes of the lattice.  Much interest has focused on possible "order-from-disorder" phenomena in this model. While the competition between spin interactions produce a massively degenerate ground state at the classical level, the degeneracy gets lifted by thermal or quantum effects, favoring a directional ordering of spin fluctuations at low temperatures \cite{Nussinov,Wenzel}. Added interest in the 2D model has been spurred by the proposal that it  describes the physics of  a collectively generated ''protected qubit'',  with a superconducting Josephson junction array as a possible realization \cite{Doucot}. Also, the model has been shown to be dual to the Xu-Moore plaquette model of $p + ip$ superconducting arrays \cite{NussinovFradkin}, and more recently, to Kitaev's toric code model in a transverse magnetic field \cite{Vidal}.

The degeneracies in the energy spectrum of the quantum compass and related orbital models make numerical simulations very demanding. In the absence of exact solutions in two or higher dimensions, progress in mapping out the phase diagram of the model is coming only  slowly and piecewise.  There is strong evidence for the existence of a symmetry-broken ground state in 2D \cite{Dorier}, however, the character of the quantum phase transition (QPT) into this state has  been a controversial issue. A continuous second-order transition is favored in one study \cite{XuMoore}, with others supporting a first-order transition \cite{Dorier,Chen,Orus}.  

Recently Brzezicki {\em et al.}~\cite{brzezicki1} considered a one-dimensional (1D) version of the quantum compass model, with the aim of analytically exploring how the interplay between quantum fluctuations and competing spin interactions produces a QPT. While the 1D model may not directly bear upon the physics of its higher-dimensional relative, it still serves as an interesting counterpoint. By a clever construction, where pairs of spin operators on neighboring sites are alternatingly mapped onto terms in a quantum Ising model (QIM), Brzezicki {\em et al.} \cite{brzezicki1} obtained an exact solution for the ground state energy,  revealing that  the 1D model exhibits a first-order transition between two disordered phases with opposite signs of certain local spin correlators.  Intriguingly, this first-order transition was found to be accompanied by a diverging correlation length for spin correlations on one sublattice. 

In this paper we exploit the method of Brzezicki {\em et al.} \cite{brzezicki1}  to study an extended version of the 1D QCM, obtained by introducing one more tunable parameter. The expanded parameter space allows us to uncover the true character of the QPT identified in Ref. \onlinecite{brzezicki1}. As follows from our analysis, this transition in fact occurs at a {\em multicritcal point} where a line of first-order transitions meets with a line of second-order transitions. The second-order critical points are Ising-like and separate locally ordered ground states from disordered states (with ordered quasi-local correlations). In contrast, when going through one of the first-order transitions, the only noticeable change in the character of the ground state is that pair correlations of spins on every second bond flip sign. In an effort to probe the anatomy of the various ground states we have analytically calculated the concurrence for pairs of spins, and also the block entanglement entropy. The concurrence signals the second-order QPTs, and the block entropy shows that they are Ising-like. None of these entanglement measures is affected by the first-order transitions.

The rest of the paper is organized as follows: In Sec.~\ref{compass} we introduce the extended 1D QCM, and apply the method of Brzezicki {\em et al.} \cite{brzezicki1} to obtain its spectrum and spin correlators. In Sec.~\ref{secQPT} we exploit our exact results to map out the zero-temperature phase diagram of the model, and identify the character of its QPTs. In Sec.~\ref{secEntanglement} we then perform an entanglement diagnostic of the various phases and phase transitions. Sec.~\ref{secConclusions}, finally, contains some concluding remarks.  

\section{The one-dimensional quantum compass model} \label{compass}
Consider the Hamiltonian
\begin{align} \label{ham}
H=\displaystyle\sum_{i=1}^{N'}\,\lbrack\,J_1\sigma^{z}_{2i-1}\sigma^{z}_{2i} + J_2\sigma^{x}_{2i-1}\sigma^{x}_{2i}
+ L_1\sigma^{z}_{2i}\sigma^{z}_{2i+1}\,\rbrack,
\end{align}
with periodic boundary conditions and where $N=2N'$ is the number of spins. In Ref.~\onlinecite{you} this is called the one-dimensional compass model, whereas the authors of Ref.~\onlinecite{brzezicki1} reserved that name for the special case with $J_1=0$ and $J_2=L_1$. It can be seen as a 1D anisotropic XY model with alternating interactions \cite{perk}. In 
Ref.~\onlinecite{brzezicki1} the model in (\ref{ham}) was studied with $J_1, J_2$ and $L_1$ 
constrained by a particular dependence on a single parameter $\alpha \in [0,1]$, with the construction mirrored in the interval $\alpha \in [1,2]$.
The solution based on this specific choice of interaction parameters indicated a first-order QPT at $\alpha=1$. However, it was not clear whether this transition is intrinsic to the system, or an artifact of the singular parameterization of the interactions. To resolve this issue, the authors of Ref.~\onlinecite{brzezicki1} revisited the problem and considered a more general model \cite{brzezicki2}, with the singular parameterization removed,
\begin{align} \label{xzhamiltonian}
H=\displaystyle\sum_{i=1}^{N'}\,\lbrack\,&J_1\sigma^{z}_{2i-1}\sigma^{z}_{2i} + J_2\sigma^{x}_{2i-1}\sigma^{x}_{2i}\nonumber \\
&+ L_1\sigma^{z}_{2i}\sigma^{z}_{2i+1} + L_2\sigma^{x}_{2i}\sigma^{x}_{2i+1}\,\rbrack,
\end{align}
with, as before, periodic boundary conditions and $N=2N'$ the number of spins. The model was diagonalized exactly by a direct Jordan-Wigner transformation, yielding an integral expression for the ground state energy in the thermodynamic limit. It was found that there is a first-order QPT when the curves $(J_1,L_2)(\alpha)$ or $(J_2,L_1)(\alpha)$, parameterized by $\alpha$, pass through $(0,0)$, thus confirming the finding in Ref.~\onlinecite{brzezicki1}. This is due to a cusp in the energy surfaces at $(J_1,L_2)=(0,0)$ and $(J_2,L_1)=(0,0)$, respectively. In fact, by carefully analyzing the energy surfaces obtained in Ref.~\onlinecite{brzezicki2} we find that there is a second-order QPT when crossing the lines $J_1=L_2$ or $J_2=L_1$, with a vanishing excitation energy gap. As these lines pass through the point $(0,0)$ of the first-order QPT, this explains the remnants of a second-order QPT that were glimpsed in Ref.~\onlinecite{brzezicki1}.

To obtain a transparent picture of the multicriticality it is actually more instructive to focus on the 1D quantum compass model in Eq.~(\ref{ham}) [identical to the Hamiltonian in (\ref{xzhamiltonian}) when $L_2=0$]. Moreover, in order to derive various spin correlation functions it is advantageous to use the approach developed in Ref.~\onlinecite{brzezicki1}. We here review this method, adapted to our case with $J_1$ and $J_2/L_1$ in Eq.~(\ref{ham}) being {\em two independent parameters}.  

We work in the conventional basis $\left\{ \left| \, \uparrow \, \rangle , \left| \, \downarrow \, \rangle \right\} \right. \right.$ of eigenstates of the $\sigma_{i}^{z}$ operators. The only terms in the Hamiltonian~(\ref{ham}) that flip spins are those which contain $\sigma^x$ operators, and these terms all have the form $\sigma_{2i-1}^x \sigma_{2i}^x$. Thus, the only transitions that can occur are simultaneous flips of spins linked by
 $\left\{2i-1,2i \right\} $ bonds (''odd bonds''). This means that the Hilbert space of the system can be divided into subspaces which are not mixed by the Hamiltonian. Each subspace can be labeled by a vector $\vec{s}=(s_1,s_2,\ldots,s_{N'})$, with element $s_i=1$ when two spins linked by an odd bond $\left\{2i-1,2i \right\} $ are parallel, and $s_i=0$ when they are antiparallel. The Hamiltonian can then be diagonalized in each subspace $\vec{s}$ independently. For any subspace $\vec{s}$ the terms involving $\sigma^{z}_{2i-1}\sigma^{z}_{2i}$ will only yield a constant contribution $C_s (J_1)$, given by
\begin{equation} \label{Cs_aniso}
C_s (J_1) = J_1\displaystyle\sum_{i\,=\,1}^{N'} \sigma^{z}_{2i-1} \sigma^{z}_{2i} = -J_1 (N'-2s),
\end{equation}
where $s \equiv \sum_{i=1}^{N'} s_i$ is the number of parallel odd bonds in the subspace $\vec{s}$.
The Hamiltonian (\ref{ham}) then becomes
\begin{equation}
H_{\vec{s}} 
= L_1 \displaystyle\sum_{i=1}^{N'}\,[\, \frac{J_2}{L_1} \,\sigma^{x}_{2i-1} \sigma^{x}_{2i} + \sigma^{z}_{2i}\sigma^{z}_{2i+1}] 
-J_1(N'-2s)\,.  \label{ovan}
\end{equation}
Next, define a set of operators 
\begin{eqnarray} 
\tau_{i}^{x} & \equiv & -\left( \, \left| \, \uparrow\downarrow \, \rangle \langle \, \downarrow\uparrow \, \right| +  \left| \, \downarrow\uparrow \, \rangle \langle \, \uparrow\downarrow  \, \right| \, \right) , \label{tauantipar1}\\
\tau_{i}^{z} & \equiv & -\left(-1\right)^{\sum_{j=1}^{i-1}s_j} \left( \, \left| \, \uparrow\downarrow \, \rangle \langle \, \uparrow\downarrow \, \right| -  \left| \, \downarrow\uparrow \, \rangle \langle \, \downarrow\uparrow \, \right| \, \right) , \label{tauantipar3}
\end{eqnarray}
for the antiparallel ($s_i=0$) odd bond $\{2i-1,2i \} $, and 
\begin{eqnarray} 
\tau_{i}^{x} & \equiv & -\left( \, \left| \, \uparrow\uparrow \, \rangle \langle \, \downarrow\downarrow \, \right| +  \left| \, \downarrow\downarrow \, \rangle \langle \, \uparrow\uparrow  \, \right| \, \right) , \label{taupar1} \\
\tau_{i}^{z} & \equiv & -\left(-1\right)^{\sum_{j=1}^{i-1}s_j} \left( \, \left| \, \uparrow\uparrow \, \rangle \langle \, \uparrow\uparrow \, \right| -  \left| \, \downarrow\downarrow \, \rangle \langle \, \downarrow\downarrow \, \right| \, \right) , \label{taupar3}
\end{eqnarray}
for the parallel ($s_i=1$) odd bond $\{2i-1,2i \}$. This maps the Hamiltonian (\ref{ham}) onto
\begin{align} \label{hamiltonian_tau}
H_{\vec{s}}=&-L_1\displaystyle\sum_{i=1}^{N'-1}[ \frac{J_2}{L_1}\tau^{x}_{i} + \tau^{z}_{i}\tau^{z}_{i+1}]\nonumber \\
&\ \ - L_1[\frac{J_2}{L_1} \tau_{N'}^{x} + (-1)^{s} \tau^{z}_{N'}\tau^{z}_{1}] + C_s(J_1)\, ,
\end{align}
which is the exactly solved \cite{LSM,barouchmccoy} one-dimensional quantum Ising model with periodic (antiperiodic) boundary conditions when $s$ is even (odd) \cite{NO}. The Hamiltonian after a Jordan-Wigner transformation to fermionic $\ci$ operators and a Fourier transformation to $\ck$ operators (see Ref.~\onlinecite{brzezicki1} for details), will then be
\begin{align}
H_{\vec{s}}^{\pm} =  L_1 \displaystyle\sum_k \, [ &\, 2((J_2/L_1)-\cos k) \ckd\ck  \nonumber \\
&+ i \sin k \,(\cmkd\ckd + \cmk \ck)\, ]  \nonumber \\
 &\qquad \quad \ \ - J_2 N' + C_{s}(J_1)\,, \label{hamiltonian_ck2}
\end{align}
where $\pm$ denote the separate subspaces of even or odd number of $c$ fermions, and $k$ takes the values $k = 0,\, \pm \frac{2\pi}{N'}, \, \pm 2\frac{2\pi}{N'}, \ldots ,\, \pi $ or $k = \pm \frac{1}{2}\frac{2\pi}{N'}, \, \pm \frac{3}{2}\frac{2\pi}{N'}, \ldots ,\, \pm\frac{1}{2}(N'-1)\frac{2\pi}{N'} $ when $s+\sum_{j=1}^{N'}\langle\cjd\cj\rangle$ is odd or even, respectively. By a Bogoliubov transformation to fermionic $\gk$ quasiparticles, the Hamiltonian in (\ref{hamiltonian_ck2}) can be expressed on diagonal form, 
\begin{equation}  \label{hamiltonian_g2}
H_{\vec{s}}^{\pm} = \displaystyle\sum_k \,  \epsilon_k \,(\gkd\gk- \frac{1}{2}) \,+\, C_{s}(J_1),
\end{equation}
where $\epsilon_k(J_2/L_1) = 2L_1(1+(J_2/L_1)^2 - 2(J_2/L_1) \cos k )^{1/2}$ and $C_{s}(J_1)=-J_1(N'-2s)$. For $s$ even and a $+$ subspace (or a $-$ subspace with $J_2/L_1 \leq 1$) the number of $\gk$ particles is even in that subspace. But for $s$ even and a $-$ subspace with $J_2/L_1 \geq 1$, the number of $\gk$ particles is odd. Similarly, for $s$ odd and a $-$ subspace (or a $+$ subspace with $J_2/L_1 \leq 1$), there must be an odd number of $\gk$ particles in that subspace. Finally, $s$ odd, + subspace, and $J_2/L_1 \geq 1$, means an even number of $\gk$ particles.

When $J_1 < 0$, the term $C_{s}(J_1)$ in Eq. (\ref{hamiltonian_g2}) forces the ground state to be in the $s$=$N'$ subspace, and not in the $s$=0 subspace as for $J_1 >0$. The ground state energy in the thermodynamic limit is then given by
\begin{equation}
E_0 \left(J_1,J_2/L_1\right) = -J_1N' -\frac{N'}{2\pi} \displaystyle\int_{-\pi}^{\pi}  dk \  \epsilon_k (J_2/L_1)
 \label{cgroundenergy}
\end{equation}
when $J_1 > 0$, and
\begin{equation}
E_0 \left( J_1, J_2/L_1 \right) = J_1N' -\frac{N'}{2\pi} \displaystyle\int_{-\pi}^{\pi}  dk \  \epsilon_k (J_2/L_1)
 \label{cgroundenergy2}
\end{equation}
when $J_1 < 0$.  Without loss of generality, we shall restrict ourselves to positive values of $J_2 $ and $L_1$ in what follows.

\section{Quantum phases and phase transitions} \label{secQPT}

\begin{figure}[t!]
	\centering
		\includegraphics[width=0.45\textwidth]{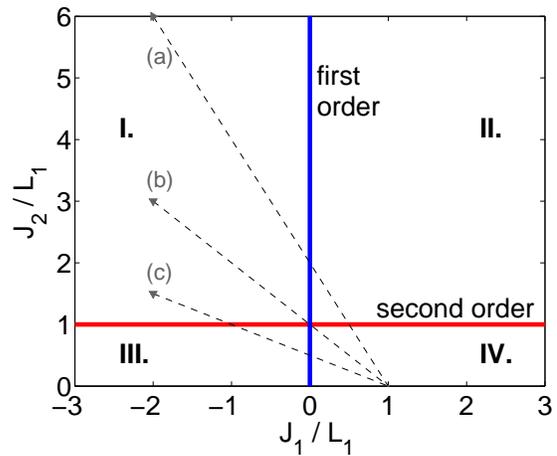}
	\caption{(Color online.) Phase diagram of the one-dimensional quantum compass model [Eq.~(\ref{ham})]. There is a first-order QPT at $J_1 = 0$ which separates the phases with  $\langle\sigma_{2i-1}^z \sigma_{2i}^z\rangle = 1$ (I. and III.) from those with  $\langle\sigma_{2i-1}^z \sigma_{2i}^z\rangle = -1$ (II. and IV.). There is also a second-order QPT at $J_2/L_1 = 1$ separating the phases with $\langle\sigma_{i}^z \rangle = 0$ (I. and II.) from those with $\langle\sigma_{i}^z \rangle \neq 0$ (III. and IV.). The dashed lines show the three paths (a), (b) and (c) that are used in Fig.~\ref{fig:gap_corrfcns}. All paths start at the point $J_1/L_1 = 1, J_2 / L_1 = 0$, where the Hamiltonian (\ref{ham}) reduces to that of the one-dimensional quantum Ising model.}
	\label{fig:phasediagram}
\end{figure}

The expressions for the ground state energy in Eqs.~(\ref{cgroundenergy}) and (\ref{cgroundenergy2}) allow us to find the QPTs of the model (\ref{ham}). There is a discontinuity in the derivative $\partial E_0 / \partial J_1$, and therefore a first-order QPT, when passing through $J_1=0$, as observed in Ref.~\onlinecite{brzezicki2}. But we also note that there is a second-order QPT when passing through $J_2 = L_1$, since the second derivatives with respect to $J_2 / L_1$ of the integrals in Eqs.~(\ref{cgroundenergy}) and (\ref{cgroundenergy2}) are divergent at $J_2 / L_1 =1$, without any singularity in first derivatives. We can thus plot the phase diagram, see Fig.~\ref{fig:phasediagram}, which shows how the lines of first- and second-order QPTs meet at the multicritical point at $J_1=0$, $J_2/L_1=1$. We now turn to investigate the different phases separated by these transitions. 
\begin{figure*}[ht!]
	\centering
		\includegraphics[width=1\textwidth]{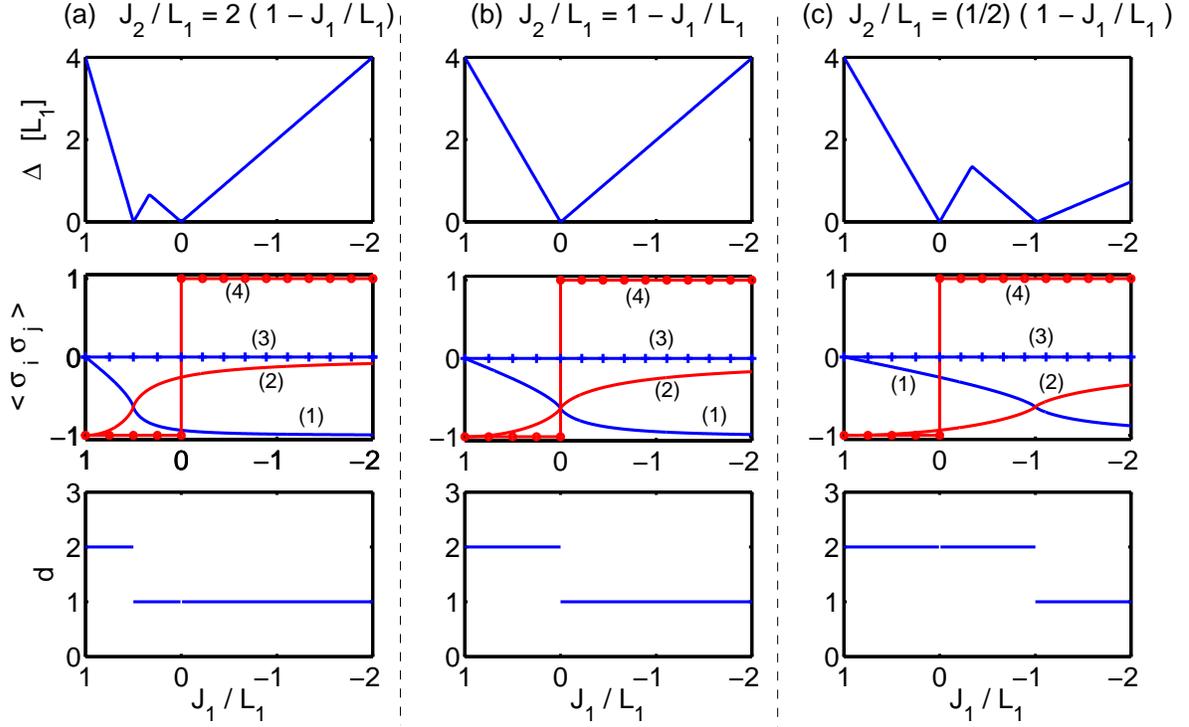}
	\caption{(Color online.) Summary of the properties of the one-dimensional compass model [Eq.~(\ref{ham})] in the thermodynamic limit. The plots are along the three paths (a) $J_2/L_1 = 2(1-J_1/L_1)$, (b) $J_2/L_1 = (1-J_1/L_1)$ and (c) $J_2/L_1 = (1/2)(1-J_1/L_1)$, shown in the phase diagram in Fig.~\ref{fig:phasediagram}. Left column: path (a). Middle column: path (b). Right column: path (c). Upper row: energy gap $\Delta$. Middle row: two-point functions in the ground state. (Blue line (1): $\langle\sigma_{2i-1}^x \sigma_{2i}^x\rangle$, red line (2): $\langle\sigma_{2i}^z \sigma_{2i+1}^z\rangle$, blue line with bars (3): $\langle\sigma_{2i}^x \sigma_{2i+1}^x\rangle$, red line with circles (4): $\langle\sigma_{2i-1}^z \sigma_{2i}^z\rangle$). Lower row: ground state degeneracy $d$.}
	\label{fig:gap_corrfcns}
\end{figure*}

Let $J_1 \neq 0$. From Eq.~(\ref{hamiltonian_ck2}) it follows that in the ground state $\langle\cpid\cpi\rangle = 0$ and $\langle\cnolld\cnoll\rangle =1$ ($\langle\cnolld\cnoll\rangle =0$) when $J_2/L_1\leq 1 $ ($J_2/L_1 > 1 $). We then find that in the thermodynamic limit, for $J_2/L_1 \leq 1$ the two states with no Bogoliubov quasiparticles, $s=0$, and either in the subspace $+$ or $-$ will be ground states, i.e. double degeneracy. For $J_2/L_1 > 1$, only the state in the $+$ subspace will be a ground state, i.e. no degeneracy. The degeneracy for $J_2/L_1 \leq 1$ will be lifted in a finite system since then the summation over $k$ in Eq.~(\ref{hamiltonian_g2}) will be different in the $\pm$ subspaces. 

When $J_1 = 0$ the energy eigenvalues are independent of $s$, since $C_s(0)=0$. Thus for a finite system, when $J_1=0$ and $J_2/L_1\neq 1$, the ground state degeneracy $d$ will be $d=2^{N'-1}=2^{N/2-1}$, since each state with no $\gamma_k$ quasiparticle in every $\{ \vec{s},+\}$ subspace with even $s$ will have the lowest energy. When $J_1=0$ and $J_2/L_1 =1$, the $\gnoll$ quasiparticle gets gapless, so that the state with one $\gnoll$ quasiparticle in every $\{ \vec{s},+\}$ subspace with odd $s$ also will have the lowest energy. Then $d$ becomes $d=2^{N'}=2^{N/2}$ for a finite system. In the thermodynamic limit each subspace $\vec{s}$ with $s$ even contributes two states to the ground state degeneracy at $J_1=0$, $J_2/L_1 =1$ (the states without $\gamma_k$ quasiparticles in both $\pm$ subspaces), and each subspace $\vec{s}$ with $s$ odd also contributes two states [the state with one $\gnoll$ quasiparticle in the $+$ subspace and the state with one $\gamma_{\pi/N'}^{\phantom{\dagger}}$ (which becomes $\gnoll$ when $N' \to \infty$) quasiparticle in the $-$ subspace]. Therefore the ground state degeneracy $d$ will be $2\times 2^{N/2}$ when $N \to \infty$. We thus confirm the result for the degeneracy at the transition point $\alpha=1$ (i.e. $J_1=0$ and $J_2/L_1 =1$) derived in Ref.~\onlinecite{brzezicki1}. The discrepancy with the result in Ref.~\onlinecite{you} where the degeneracy was found to be $d=2^{N/2-1}$ remains unexplained. 

The energy gap between the ground state and the lowest excitations is obtained from the diagonalized Hamiltonian in Eq.~(\ref{hamiltonian_g2}) and the subsequent rules for the number of $\gk$ particles allowed in each subspace. In the thermodynamic limit, the gaps are $\Delta_1 = 4L_1|1-J_2/L_1|$ for the state in the $s=0$ subspace with one $\gnoll$ and one $ \gamma_{\pm2\pi/N'}^{\phantom{\dagger}}$ particle, $\Delta_2 = 2|J_1|$ for the state in the $s=1$ subspace with no $\gk$ particles (this is only possible when $J_2/L_1 \geq 1$), $\Delta_3 = 4|J_1|$ for the state in the $s=2$ subspace with no $\gk$ particles, and $\Delta_4 = 2L_1|1-J_2/L_1|$ for the state in the $s=0$ subspace with one $\gnoll$ particle (which is only possible when $J_2 / L_1 \geq 1$). At each point in the phase diagram the excitation energy gap from the ground state is now given by the smallest of $\Delta_{1,2,3,4}$. This gap is plotted in Fig.~\ref{fig:gap_corrfcns}, along three representative lines in the phase diagram (cf. Fig.~\ref{fig:phasediagram}). The first- and second-order QPTs are marked by vanishing energy gaps.

As previously seen, the ground state is in the $s$=$N'$ subspace when $J_1 < 0$, and in the $s$=0 subspace when $J_1 >0$. Therefore the ground state two-point functions $\langle \sigma_{2i-1}^{z} \sigma_{2i}^{z} \rangle$ are discontinuous at $J_1 = 0$, with $\langle \sigma_{2i-1}^{z} \sigma_{2i}^{z} \rangle =1$ when  $J_1 < 0$ and $\langle \sigma_{2i-1}^{z} \sigma_{2i}^{z} \rangle =-1$ when  $J_1 > 0$. On the other hand, $\langle \sigma_{2i}^{x} \sigma_{2i+1}^{x} \rangle=0$ in all phases, since the operator $ \sigma_{2i}^{x} \sigma_{2i+1}^{x} $ takes the ground state out of its subspace $\vec{s}$. The mapping to the QIM [Eqs.~(\ref{tauantipar1}) to (\ref{taupar3})] implies that $\sigma^{x}_{2i-1}\sigma^{x}_{2i} \mapsto - \tau_{i}^x$ and $\sigma^{z}_{2i}\sigma^{z}_{2i+1} \mapsto -\tau_{i}^{z} \tau_{i+1}^{z}$, so that $\langle \sigma^{x}_{2i-1}\sigma^{x}_{2i} \rangle =  -\langle \tau_{i}^x \rangle$ and $\langle \sigma^{z}_{2i}\sigma^{z}_{2i+1} \rangle=  -\langle \tau_{i}^{z} \tau_{i+1}^{z}\rangle$.

The characteristics of the different phases and phase transitions of the model are graphically summarized in Fig.~\ref{fig:gap_corrfcns}, where we plot the energy gap, the two-point functions and the ground state degeneracy along the three paths (a), (b) and (c) shown in Fig.~\ref{fig:phasediagram}. It is seen in Fig.~\ref{fig:gap_corrfcns} that the continuous QPT at $J_2/L_1 =1$ is associated with a continuous transition from dominating antiparallel ordering of spin $z$ components on even bonds for $J_2/L_1< 1$, to dominating antiparallel ordering of spin $x$ components on odd bonds for $J_2/L_1 > 1$. The order parameter $\langle \sigma^z_{2i} \rangle$ or $\langle \sigma^z_{2i-1} \rangle$ is only non-zero in the ordered phase given by $J_2/L_1 < 1$, since $|\langle \sigma^z_{2i} \rangle| = |\langle \sigma^z_{2i-1} \rangle| = \langle \tau^z_{i} \rangle$. The first-order QPT at $J_1=0$ corresponds to $\alpha = 1$ in Ref.~\onlinecite{brzezicki1}. We see that the mixed first- and second-order features of the QPT in the one-parameter model studied by the authors of Ref.~\onlinecite{brzezicki1} come about because of their parameterization, where the multicritical point is approached along a path [(b) in Fig.~\ref{fig:phasediagram}] with projections along both of the transition lines.  

At $J_1=0$ the Hamiltonian (\ref{ham}) reduces to what Ref.~\onlinecite{brzezicki1} refers to as the proper one-dimensional QCM,
\begin{equation} \label{compasshamiltonian2}
H=\displaystyle\sum_{i=1}^{N'}\,\lbrack\, J_2\sigma^{x}_{2i-1}\sigma^{x}_{2i}
+ L_1\sigma^{z}_{2i}\sigma^{z}_{2i+1}\,\rbrack.
\end{equation}
It was shown above that the ground state degeneracy for a finite system is $d=2^{N/2-1}$, except at the second-order QPT at $J_2=L_1$, where it is $d=2^{N/2}$. The large ground state degeneracy can thus be easily understood as a consequence of the model (\ref{compasshamiltonian2}) being at the level crossings of the first-order QPT of the extended 1D QCM~(\ref{ham}).

\section{Entanglement} \label{secEntanglement}
Given the exact solution of the 1D QCM, we have a rare opportunity to analytically probe for the entanglement in the ground state of a highly complex system of coupled qubits. We here focus on two of the most frequently used entanglement measures: {\em concurrence} and {\em block entanglement entropy} \cite{amico}.

\subsection{Concurrence}
The concurrence of two spins at sites $i$ and $j$ is obtained from their reduced density matrix $\rho_{ij}$, which can be expanded as~\cite{osbornenielsen}
\begin{equation} \label{rdm2}
\rho_{ij} = \frac{1}{4} \displaystyle \sum_{\mu,\nu} \langle \sigma_i^{\mu} \sigma_j^{\nu}\rangle \sigma_i^{\mu} \sigma_j^{\nu}.
\end{equation} 
where $\mu,\nu=0,x,y,z$, with $\hat{\sigma}_i^{0} \equiv \openone_i$. Note that the ground state expectation values of $\sigma_i^{x}$, $\sigma_i^{y}$, $ \sigma_{i}^{x}\sigma_{j}^{z} $ and $ \sigma_{i}^{y}\sigma_{j}^{z} $ (and $i \leftrightarrow j$) are zero, since these operators flip only one spin on an odd bond, giving a state in a different subspace than the ground state. The correlation function $\langle \sigma_{i}^{x}\sigma_{j}^{y} \rangle$ must also be zero, since the matrix $\sigma_{i}^{x}\sigma_{j}^{y}$ is imaginary and $\rho_{ij}$ must be real as the Hamiltonian is real. Therefore, the reduced density matrix (\ref{rdm2}) becomes
\begin{align}  \label{rdm22}
\rho_{ij} = \frac{1}{4} (1 &+ \langle \sigma_{i}^{z}\rangle  \sigma_{i}^{z} + \langle \sigma_{j}^{z}\rangle  \sigma_{j}^{z}   + \langle \sigma_{i}^{x}\sigma_{j}^{x} \rangle  \sigma_{i}^{x}\sigma_{j}^{x} \nonumber \\
&+ \langle \sigma_{i}^{y}\sigma_{j}^{y} \rangle \sigma_{i}^{y}\sigma_{j}^{y} + \langle \sigma_{i}^{z}\sigma_{j}^{z} \rangle \, \sigma_{i}^{z}\sigma_{j}^{z} ) \,.
\end{align}

The concurrence $C(\rho_{ij})$ is now given by~\cite{hill} $C(\rho_{ij})=\max \left\{\,0\,,\,\lambda_1 - \lambda_2 - \lambda_3 - \lambda_4 \,\right\}$, where $\lambda_1 \ge \lambda_2 \ge \lambda_3 \ge \lambda_4$ are the non-negative real eigenvalues of the Hermitian matrix $R \equiv \sqrt{\sqrt{\rho_{ij}}\tilde{\rho}_{ij}\sqrt{\rho_{ij}}}$. In this expression, $\tilde{\rho}_{ij}=(\sigma^{y}_i  \sigma^{y}_j)\,\rho^{\ast}_{ij}\,(\sigma^{y}_i\sigma^{y}_j)$, where $\rho^{\ast}_{ij}$ is the complex conjugate of $\rho_{ij}$ in the given basis. Here, the eigenvalues $r_{1,2,3,4}$ of $R$ are (without ordering)
\begin{eqnarray} 
r_{1,2}& = &\frac{1}{4}\, | \, \sqrt{1+ \langle \sigma_{i}^{z}\rangle + \langle \sigma_{j}^{z}\rangle + \langle \sigma_{i}^{z}\sigma_{j}^{z}\rangle} \nonumber \\
&& \quad \times \sqrt{1- \langle \sigma_{i}^{z}\rangle - \langle \sigma_{j}^{z}\rangle + \langle \sigma_{i}^{z}\sigma_{j}^{z}\rangle} \nonumber \\
&& \quad \pm | \langle \sigma_{i}^{x}\sigma_{j}^{x}\rangle - \langle \sigma_{i}^{y}\sigma_{j}^{y}\rangle |\, |, \label{eig1} \\
r_{3,4} &=& \frac{1}{4}\, | \, \sqrt{1+ \langle \sigma_{i}^{z}\rangle - \langle \sigma_{j}^{z}\rangle - \langle \sigma_{i}^{z}\sigma_{j}^{z}\rangle} \nonumber \\
&& \quad \times \sqrt{1- \langle \sigma_{i}^{z}\rangle + \langle \sigma_{j}^{z}\rangle - \langle \sigma_{i}^{z}\sigma_{j}^{z}\rangle}\nonumber \\
&& \quad \pm | \langle \sigma_{i}^{x}\sigma_{j}^{x}\rangle + \langle \sigma_{i}^{y}\sigma_{j}^{y}\rangle |\, |.\label{eig4}
\end{eqnarray}
It is immediately clear that the concurrence of two spins that are not on the same odd bond is zero, since then $\langle \sigma_{i}^{x}\sigma_{j}^{x}\rangle = \langle \sigma_{i}^{y}\sigma_{j}^{y}\rangle = 0$ which gives $r_1=r_2$ and $r_3=r_4$. For two spins that are on the same odd bond $\{2i-1,2i\}$, we have that $\langle \sigma_{2i-1}^{x} \sigma_{2i}^{x}\rangle =- \langle\tau_{i}^x \rangle$. When $J_1 > 0$, the ground state is in the subspace $s$=$0$, so then $\langle \sigma_{2i-1}^{z}\rangle = - \langle \sigma_{2i}^{z}\rangle=-\langle\tau_{i}^z \rangle$ and $\langle \sigma_{2i-1}^{z} \sigma_{2i}^{z}\rangle = -1$. We also have that $\langle \sigma_{2i-1}^{y}\sigma_{2i}^{y} \rangle = \langle \sigma_{2i-1}^{x}\sigma_{2i}^{x} \rangle$, since the matrix $ \sigma_{2i-1}^{y}\sigma_{2i}^{y} $ is the same as $ \sigma_{2i-1}^{x}\sigma_{2i}^{x} $ in the subspace $s$=$0$. Then $C(\rho_{2i-1,2i})=  \max \{0\,,\,r_3 - r_4\} = -\langle \sigma_{2i-1}^{x}\sigma_{2i}^{x}\rangle=\langle\tau_{i}^x \rangle$, since $\sqrt{1-\langle\tau_{i}^z \rangle^2} \geq \langle\tau_{i}^x \rangle$. On the other hand, when $J_1 < 0$, the ground state is in the subspace $s$=$N'$, where $\langle \sigma_{2i-1}^{x} \sigma_{2i}^{x}\rangle =- \langle\tau_{i}^x \rangle$, $\langle \sigma_{2i-1}^{z}\rangle =  \langle \sigma_{2i}^{z}\rangle=-\langle\tau_{i}^z \rangle$, and $\langle \sigma_{2i-1}^{z} \sigma_{2i}^{z}\rangle = 1$. Also, in the subspace $s$=$N'$ the matrix $ \sigma_{2i-1}^{y}\sigma_{2i}^{y} $ is the same as $ -\sigma_{2i-1}^{x}\sigma_{2i}^{x} $, so that now $\langle \sigma_{2i-1}^{y}\sigma_{2i}^{y} \rangle = -\langle \sigma_{2i-1}^{x}\sigma_{2i}^{x} \rangle$. We then get the result that $C(\rho_{2i-1,2i})=  \max \{0\,,\,r_1 - r_2\} = -\langle \sigma_{2i-1}^{x}\sigma_{2i}^{x}\rangle=\langle\tau_{i}^x \rangle$. Thus, for all values of $J_1$, the only pairs of spins with non-zero concurrence are those on the same odd bond, and for them $C(\rho_{2i-1,2i})=  \langle\tau_{i}^x \rangle$. The pairwise concurrence therefore only depends on the parameter $J_2 / L_1$, which plays the role of the magnetic field in the QIM (\ref{hamiltonian_tau}). From the well-known solution \cite{LSM,barouchmccoy} of the 1D QIM, it follows that in the thermodynamic limit the concurrence has a diverging first derivative across the curve $J_2/L_1= 1$ of the second-order QPT. Note that the first-order QPT at $J_1 = 0$ is not signaled by the pairwise concurrence. In particular, this means that when going through the multicritical point, the concurrence will behave as if it was a pure second-order transition, despite the transition actually being of first order.

\subsection{Block entropy}

The block entropy is the von Neumann entropy of a subsystem consisting of an entire block of adjacent spins, thus giving a measure of the amount of entanglement between the block and the rest of the system \cite{amico}. We will now show that a block of an even number of $\sigma$ spins in the 1D QCM~(\ref{ham}) that fully covers an integer number of odd bonds will have the same block entanglement as half the number of $\tau$ spins in the corresponding QIM (\ref{hamiltonian_tau}). Let us first consider the case when $J_1>0$. 

The reduced density matrix $\rho_L$ for a block of $L$ spins $\sigma_1, \ldots,\sigma_L$, where $L$ is even and $\sigma_1$ and $\sigma_2$ are on the same odd bond, can be expanded as \cite{osbornenielsen}
\begin{equation} \label{reddensL}
\rho_{L} = \frac{1}{2^L} \displaystyle\sum_{\mu_1,\ldots,\mu_L} \langle \sigma_{1}^{\mu_1}\ldots \sigma_{L}^{\mu_L} \rangle \, \sigma_{1}^{\mu_1}\ldots \sigma_{L}^{\mu_L}\,,
\end{equation}
where $\mu_1,\ldots,\mu_L$ are summed over $0,x,y,z$. For the expectation value $\langle \sigma_{1}^{\mu_1}\ldots \sigma_{L}^{\mu_L} \rangle$ not to be zero, the only allowed pairs of $\sigma$ operators on every odd bond in the summation are
\begin{align*}
\sigma_{2i-1}^0 \sigma_{2i}^0  &\mapsto  \tau_i^0 ,&
\sigma_{2i-1}^z \sigma_{2i}^z  &\mapsto  -\tau_i^0 , \\
\sigma_{2i-1}^z \sigma_{2i}^0  &\mapsto   -\tau_i^z, & 
\sigma_{2i-1}^0 \sigma_{2i}^z  &\mapsto   \tau_i^z , \\
\sigma_{2i-1}^x \sigma_{2i}^x  &\mapsto   -\tau_i^x ,& 
\sigma_{2i-1}^y \sigma_{2i}^y  &\mapsto  -\tau_i^x , \\
\sigma_{2i-1}^y \sigma_{2i}^x  &\mapsto  \tau_i^y , &
\sigma_{2i-1}^x \sigma_{2i}^y  &\mapsto  -\tau_i^y  .
\end{align*}
Thus the reduced density matrix (\ref{reddensL}) becomes
\begin{equation}
\rho_L = \frac{1}{2^{L/2}}  \displaystyle\sum_{\nu_1,\ldots,\nu_{L/2}} \langle \tau_1^{\nu_1} \ldots \tau_{L/2}^{\nu_{L/2}} \rangle \, \tau_1^{\nu_1} \ldots \tau_{L/2}^{\nu_{L/2}} \,,
\end{equation}
where $\nu_1,\ldots,\nu_{L/2}$ are summed over $0,x,y,z$. This is precisely the reduced density matrix of $L/2$ spins in the QIM. The block entropy $S_L$ is then the same as the block entropy of $L/2$ spins in the QIM with effective transverse field equal to $J_2/L_1 $. This will apply equally well also when $J_1<0$.  The phase transition at $J_2/L_1  = 1$ is therefore in the Ising universality class, since this is uniquely determined by the scaling of the block entropy at criticality \cite{CalabreseCardy}. The 
first-order QPT at $J_1=0$ and with $J_2/L_1  \neq 1$, corresponds to a non-critical QIM, with saturated block entropy $S_L$ when $L \to \infty$. Its value will be the same whether $J_1 \to 0^-$ or $J_1 \to 0^+$. 

\section{Conclusions} \label{secConclusions}
To summarize, we have performed an exact analytical study of the 1D quantum compass model, using a mapping to the quantum Ising model, following the approach in Ref.  \onlinecite{brzezicki1}. 
We identify four distinct ground state phases, separated by two intersecting transition lines. One of these defines a line of second-order Ising-like transitions, while the other is a line of first-order transitions (cf. Fig.~\ref{fig:phasediagram}).  The point of intersection, where the first-order quantum phase transition identified by Brzezicki {\em et al.} \cite{brzezicki1} takes place, thus defines a {\em multicritical point}. This explains the apparently exotic behavior at the transition found by these authors. In particular, the appearance of a diverging correlation length for certain spin correlations finds a natural explanation once the multicriticality of the transition point has been recognized. One may ask whether the quantum phase transition in the 2D quantum compass model maybe also plays out at a multicritical point, in analogy to the 1D model? If so, this could possibly explain the notorious difficulty in identifying the character of the transition, as evidenced by the conflicting results in Ref. \onlinecite{XuMoore} and Refs. \onlinecite{Dorier,Chen,Orus}.

Our results for the entanglement show that the only effect on the ground state when going through the first-order transitions is that a correlation function for
neighboring spins on odd bonds changes sign, without any effect on the entanglement measures studied. First-order QPTs are generally associated with a discontinuity in concurrence, but ''accidental'' exceptions to this rule are possible \cite{wu}. We here have an example thereof.

{\bf Acknowledgments.} \ The authors wish to thank Jun-Peng Cao, Shi-Jian Gu, Simon Trebst, and Stellan \"Ostlund for helpful discussions. We also acknowledge the Kavli Institute for Theoretical Physics at UCSB for hospitality during the completion of this work. This research was supported in part by the National Science Foundation under Grant No. PHY05-51164, and by the Swedish Research Council under Grant No. VR-2005-3942.


\begin{thebibliography}{99}

\bibitem{Cheong}
For a brief survey, see S.-W. Cheong, Nature Mater. {\bf 6}, 927 (2007).  

\bibitem{KugelKhomskii}
K. I. Kugel and D. I. Khomskii, Sov. Phys. JETP {\bf 37}, 725 (1973).

\bibitem{Nussinov}
Z. Nussinov, M. Biskup, L. Chaves, and J. van den Brink, Europhys. Lett. {\bf 67}, 990 (2004).

\bibitem{Wenzel}
S. Wenzel and W. Janke, Phys. Rev. B {\bf 78}, 064402 (2008). 

\bibitem{Doucot}
B. Dou\ifmmode \mbox{\c{c}}\else \c{c}\fi{}ot, M. V. Feigel\char39{}man, L. B. Ioffe, and A. S. Ioselevich, Phys. Rev. B {\bf 71}, 024505 (2005).

\bibitem{NussinovFradkin}
Z. Nussinov and E. Fradkin, Phys. Rev. B {\bf 71}, 195120 (2005).

\bibitem{Vidal}
J. Vidal, R. Thomale, K. P. Schmidt, and S. Dusuel, e-print arXiv:0902.3547.

\bibitem{Dorier}
J. Dorier, F. Becca, and F. Mila, Phys. Rev. B {\bf 72}, 024448 (2005).

\bibitem{XuMoore}
C. Xu and J. E. Moore, Phys. Rev. Lett. {\bf 93}, 047003 (2004).

\bibitem{Chen}
H.-D. Chen, C. Fang, J. Hu, and H. Yao, Phys. Rev. B {\bf 75}, 144401 (2007).

\bibitem{Orus}
R. Or\'{u}s, A. C. Doherty, and G. Vidal, Phys. Rev. Lett. {\bf 102}, 077203 (2009).

\bibitem{brzezicki1}
W. Brzezicki, J. Dziarmaga, and A.~M. Ole\'{s}, Phys. Rev. B {\bf 75},  134415 (2007).
  
\bibitem{you}
W.-L. You and G.-S. Tian, Phys. Rev. B {\bf 78},  184406  (2008).

\bibitem{perk}
J.~H.~H. Perk and H.~W. Capel, Physica A {\bf 89}, 265 (1977).

\bibitem{brzezicki2}
W. Brzezicki and A.~M. Ole\'{s}, Acta Phys. Pol. A {\bf 115}, 162 (2009) (e-print arXiv:0805.3904).

\bibitem{LSM}
E. Lieb, T. Schultz, and D. Mattis, Ann. Phys. (N.Y.) {\bf 16},  407  (1961); S. Katsura, Phys. Rev. {\bf 127}, 1508 (1962).

\bibitem{barouchmccoy}
P. Pfeuty, Ann. Phys. (N.Y.) {\bf 57}, 79 (1970); E. Barouch and B.~M. McCoy, Phys. Rev. A {\bf 3},  786  (1971).

\bibitem{NO}
For a ''bond algebraic'' version of the mapping, see Z. Nussinov and
G. Ortiz, e-print arXiv:0812.4309.  

\bibitem{amico}
For a review, see L. Amico, R. Fazio, A. Osterloh, and V. Vedral, Rev. Mod. Phys. {\bf 80},  517
  (2008).
 
\bibitem{osbornenielsen}
T.~J. Osborne and M.~A. Nielsen, Phys. Rev. A {\bf 66},  032110  (2002).

\bibitem{hill}
S. Hill and W.~K. Wootters, Phys. Rev. Lett. {\bf 78},  5022  (1997).

\bibitem{CalabreseCardy}
P. Calabrese and J. Cardy, J. Stat. Mech.: Theory Exp. 2004, P06002 (2004).

\bibitem{wu}
L.-A. Wu, M.~S. Sarandy and D.~A. Lidar, Phys. Rev. Lett. {\bf 93}, 250404 (2004).



\end{thebibliography}
\end{document}